# Tamm plasmons in metal/nanoporous GaN distributed Bragg reflector cavities for active and passive optoelectronics

Guillaume Lheureux[1,2,†,*], Morteza Monavarian[1,2,†,*], Ryan Anderson[1,2,*], Ryan A. DeCrescent[3], Joel Bellessa[4], Clémentine Symonds[4], Jon A. Schuller[5], Shuji Nakamura[1,2], James S. Speck[1,2], and Steven P. DenBaars[1,2]

[1]*Materials Department, University of California Santa Barbara, Santa Barbara, California 93106, USA*

[2]*Solid-State Lighting Energy Electronic Center (SSLEEC), University of California Santa Barbara, Santa Barbara, California 93106, USA*

[3]*Department of Physics, University of California Santa Barbara, Santa Barbara, California 93106, USA*

[4]*Université Lyon, Université Claude Bernard Lyon 1, CNRS, Institut Lumière Matière, F-69622, Lyon, France*

[5] *Department of Electrical and Computer Engineering, University of California Santa Barbara, Santa Barbara, California 93106, USA*

[†]*These authors contributed equally to this work.* *[glheureux@ucsb.edu](mailto:glheureux@ucsb.edu) , [mmonavarian@ucsb.edu](mailto:mmonavarian@ucsb.edu)*

We investigate Tamm plasmon (TP) modes in a metal/semiconductor distributed Bragg reflector (DBR) interface. A thin Ag (silver) layer with an optimized thickness (~ 55 nm from simulation) was deposited on nanoporous GaN DBRs fabricated using electrochemical etching on freestanding semipolar ($20\bar{2}1$) GaN substrates. The reflectivity spectra of the DBRs are compared before and after the Ag deposition and with that of a blanket Ag layer deposited on GaN. The results indicate presence of a TP mode at ~ 455 nm on the structure after the Ag deposition. An active medium can also be accommodated within the mode for optoelectronics and photonics. Moreover, the simulation results predict a sensitivity of the TP mode wavelength to the ambient (~ 4 nm shift when changing the ambient within the pores from air with $n = 1$ to isopropanol $n = 1.3$), suggesting an application of the nanoporous GaN based TP structure for optical sensing.

A Tamm plasmon (TP) mode is an optical state that exists at the interface between a metallic layer and a distributed Bragg reflector (DBR). Named in analogy to the Tamm state[1], a class of defect state at the interface of crystal, TP modes were first experimentally demonstrated by Sasin *et al.* in 2008[2] shortly after its theoretical prediction in 2006-2007 [3,4]. TP modes behave as "zero dimensional" (0D) cavity modes with a strong vertical confinement, resulting in a high quality factor over 1000[5]. In contrast to other optical surface state associated with metals, TP mode can be either TE- or TM-polarized and their dispersion relations lies partially within the light cone. Hence, the TP modes can be directly coupled to light modes or surface plasmons[6], which makes them extremely interesting for optoelectronic applications.

TP-based Lasers[7–9], single photon emitters[10] and sensors[11] have already been demonstrated in the past few years. One of their most interesting benefit resides in that the optical properties of TP modes can be largely tuned by tailoring the metallic part without the need to modify the complete structure[12,13]. Historically used in Arsenide material systems, TP structures have recently also been introduced in III-nitride systems mainly to improve the light-extraction efficiency (LEE) of light-emitting diodes (LEDs) emitting in green[14]. TP modes have also been realized in dielectric mesoporous DBRs[11], while the dielectric DBRs cannot be easily integrated into semiconductor based optoelectronics and photonics platforms.

Various material systems can be used for DBR mirrors in the TP structures, including dielectric and semiconductor materials. Dielectric DBRs typically provide high refractive index contrast between the layers, while may not be easily integrated into semiconductor technology due to the epitaxial difficulties [15]. On the other hand, semiconductor DBRs suffer from low refractive index contrast and typically large lattice mismatch between the layers, particularly for nitride systems[16]. Ternary strain-compensated InAlN lattice matched to GaN have been used as DBRs for lasers[17]. However, the growth of strain-compensated ternary DBRs requires much more complicated growth conditions and suffers from a small growth window. Nanoporous GaN technology have been proposed as an alternative to other DBRs, which provides perfectly lattice matched condition, with refractive index contrasts as large as those for dielectric DBR systems[18,19]. In addition, the porosity of the nanoporous GaN in the DBRs can be tuned by the doping and EC etching conditions[19] (the size and density of the pores are directly correlated with the bias voltage and doping density, respectively[18]). Therefore, the technology may provide access to a long range of wavelengths for various applications[20–22]. At the contrary to dielectric mesoporous DBR, the nanoporous GaN DBR allows for the realization of active devices with an easy integration in the industrial development process.

In this work, we demonstrate the first observation of a TP mode appearing at the interface between nanoporous GaN DBR and a thin silver (Ag) layer. First, we theoretically calculate the position of the TP modes indicating a strong dependence of TP mode on the thickness of Ag layer as well as the porosity of the DBR layers. A steep dip in the reflectivity spectrum of the DBR after depositing the Ag layer clearly indicates the presence of a TP mode at ~ 455 nm. Our simulation results indicate a ~ 4 nm shift of the TP mode position when the nanoporous layer refractive index changes from 1.0 to 1.3, indicating a strong potential of such structures for optical sensing applications. The nanoporous GaN based TP modes can be easily integrated into optoelectronic and photonic technologies. The results suggest using nanoporous GaN based TP structures for active (*e.g.* light-emitting diodes, lasers) and passive (*e.g.* optical sensors) structures and beyond.

Optical simulation of TP structure and DBR were carried out using a transfer-matrix algorithm. The optical index of the silver layer is obtained using a polynomial fitting of the experimental values obtained by Johnson and Christies[23]. The optical index $n_{low}$ of the porous layer is given by the volume average theory (VAT) as[19]

$$n_{low} = \sqrt{(1-\rho)n_{GaN}^2 + \rho\, n_{por}^2} \qquad (1)$$

Where $\rho$ is the porosity of the layer, $n_{GaN}$ is the refractive index of GaN (chosen here as 2.36[24]) and $n_{por}$ is the refractive index of the material inside the pore, normally considered as $n_{por} = 1$ for air. In order to take into account the experimental fluctuations in the porosity that may occur during the fabrication of the nanoporous DBRs, the porosity of the low index layers were randomly chose following a normal law centered around 53% of porosity with a fluctuation of 10%. The applied fluctuations did not show a drastic impact on the final results. The central wavelength $\lambda_0$ of the GaN/GaN DBR was chosen equal to 415 nm.

Figure 1(a) shows the calculated reflectivity spectrum of a nanoporous GaN 16 pairs GaN/GaN DBR with (solid red line) and without (dash black line) 55 nm of Ag. The calculated dispersion relation of the two different structures are shown in Figure 1(b). The TP mode appears as a sharp resonance within the stop band of the DBR. It's interesting to notice that this is the first reflectivity minimum (rebound) of the DBR, at the stop band edge, that give rise to the TP mode, as showed in figure 1(c) which display the reflectivity spectra of a TP structure as a function of the silver layer thickness. A blue shift of the resonance mode and a reduction of the resonance linewidth with increasing the Ag layer thickness are apparent (Figure 1(c)). The linewidth (or the quality factor) of the TP mode in the reflectivity of reaches an optimum value at ~ 50-55 nm of Ag. Figures 1(d) and (e) indicate the tunability of the TP mode by adjusting $n_{por}$ and $\rho$, respectively. Keeping the layer thickness constant, an increase either in the value of $n_{por}$ or $\rho$ lead to a red shift of the TP mode. By increasing the porosity $\rho$, the optical constrast between the low index and high index layer is increased and it lead to a larger stop band which push the TP mode to the higher wavelength. On the other side, modifying the optical index of the low index layer by changing the material in the pores should lead to a blue shift of the TP mode because of the reduced stop band compared to the ambient. But because we are keeping a constant thickness $d$ of our layers ($\lambda_0/4n_{GaN}$ and $\lambda_0/4n_{low}$ for $\rho = 53\%$), the central wavelength of the DBR is also redshifted as the actual thickness of the layer are not matching the quarter wavelength rules and the TP mode is also redshifted compared to the ambient.

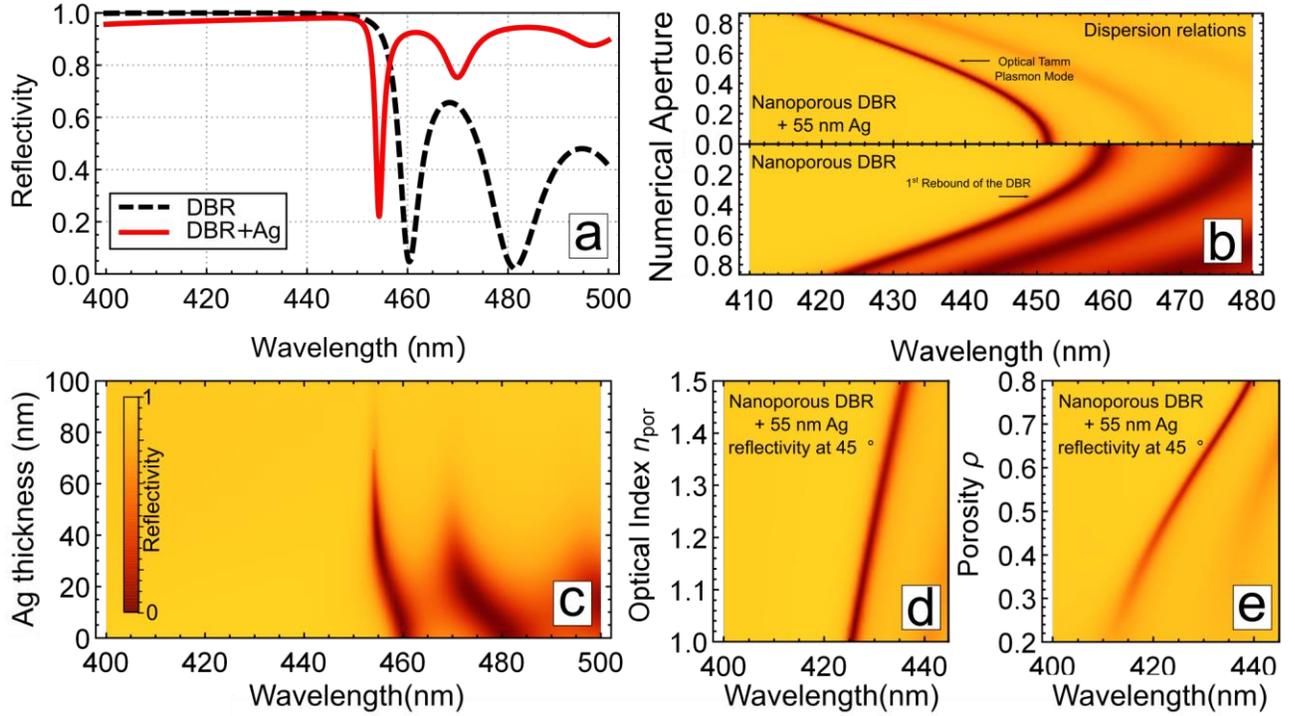

Figure 1. (a) Simulated reflectivity of a 16 pairs nanoporous GaN DBR (dash black line) and a 55 nm silver layer on top of the nanoporous GaN DBR (solid red line). (b) TM Simulated dispersion relation of Tamm plasmon mode (top) and a DBR (bottom). (c) Calculated reflectivity spectra of TP structure as a function of the silver thickness layer. (d) Calculated reflectivity spectra of TP structure at 45° incidence angle as a function of $n_{por}$ in the low-index nanoporous GaN layer. (e) Calculated reflectivity spectra of TP structure at 45° incidence angle as a function of ρ in the low-index nanoporous GaN layer.

To experimentally verify the simulation results, a TP structure composed of a 55 nm Ag layer (designed based on the optimum thickness from simulation result of Figure 1(c)) on top of a nanoporous GaN DBR was fabricated. The DBR sample was grown on semipolar $(20\bar{2}\bar{1})$ bulk GaN substrate from Mitsubishi Chemical Corporation (MCC) using an atmospheric vertical design metalorganic chemical vapor deposition (MOCVD), where trimethylgallium (TMGa), $NH_3$, and $SiH_4$ were used as sources of elemental Ga, N, and Si dopants, respectively. A 300 nm-thick unintentionally doped (UID) GaN buffer layer was first grown on the substrate to improve the morphology. Then, a periodic 16 pairs of alternating doped and UID GaN layers were grown with the designed thicknesses of 60 nm and 45 nm, respectively to target the desired stop-band wavelengths. The doping level for the *n*-type layer was [Si]: $5.5 \times 10^{19}$ cm$^{-3}$, as verified by secondary-ion mass spectroscopy (SIMS). All layers were grown at 1180 °C and atmospheric pressure.

The DBR was then formed using a wet electrochemical (EC) etching technique, as shown in Figure 2. Rectangular trenches were etched down to the substrate using standard photolithography and a $Cl_2$-based reactive ion etching (RIE) opening up a window to the highly doped layers. An indium metal contact was rubbed on the backside of the sample and held with conductive forceps, partially submerging the sample in an oxalic acid bath while keeping the metal contact away from the solution (Figure 2(a)). The highly doped layers were then porosified by applying a voltage through a submerged Pt mesh cathode at 4V for 2 hours and 200 rpm stirring. A strong dependence of the EC porosification process on the bias voltage and the doping level is expected[18,19], as schematically shown in Figure 2(b). The porous GaN area was later verified to undercut 30 µm in from the trench edges at approximately 50% porosity determined by focused ion beam (FIB) cross-sectional and SEM imaging. Figure 3 shows a high-resolution cross-sectional scanning electron microscopy (SEM) image of the processed DBR after EC etching. All the doped layers were uniformly porosified resulting in pore sizes ~ 30-50 nm. A porosity of ~ 50% was realized from evaluation of the SEM image of Figure 3 using an approach similar to what is shown in Ref. 18. A refractive index of $n_{low}$~1.83 was realized for the porous layer by the VAT, as described above (Eq. (1)). Reflectance spectra of the structure were recorded using a commercial Thin-Film UV visible spectrometers (Filmetrics F10-RT-UVX). A 55 nm Ag layer was then blanket deposited on the DBR sample using an electron-beam evaporator.

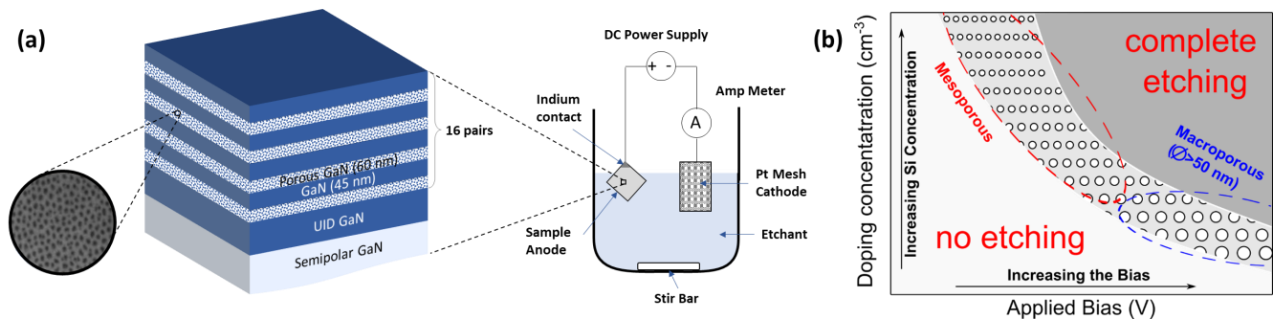

Figure 2. (a) Schematics of an electrochemical etching setup used to porosify the doped GaN layers and form the nanoporous GaN DBRs. (b) Schematic representation of different etching regimes in the electrochemical etching of GaN, indicating the strong effect of applied bias and doping concentration on the porosity of the nanoporous GaN (from no etching regime to mesoporous/macroporous, and to complete etching regime).

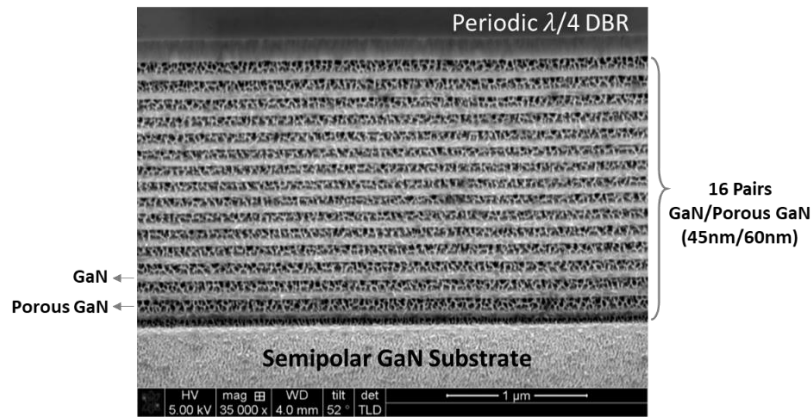

Figure 3. High-resolution SEM image of a 16-pair nanoporous GaN DBR processed using the EC etching method.

Figure 4(a) shows an experimentally measured reflectivity spectra of the DBRs before and after Ag deposition. According to the plot of Figure 4(a), the DBR reflectivity spectra shows a stop-band below 450 nm (dash black line in Figure 4(a)). The first rebound of the DBR at 465 nm is not well visible because of potential nonuniformity of the porosification within the measurement spot size, which are not detrimental to the stop band[25] The spectrum of the DBR structure is measured at the same spot before and after the Ag blanket deposition. The reflectivity of the TP structure (DBR + 55 nm Ag) shows a steep dip near 456 nm (solid red line in Figure 4(a)) indicating a TP mode, ~ 10 nm below the first rebound wavelength of the DBR, which is in good agreement with the simulation results of Figure 1(a). The spectrum of the Ag + GaN is also plotted (grey dots in Figure 4(a)) for comparison.

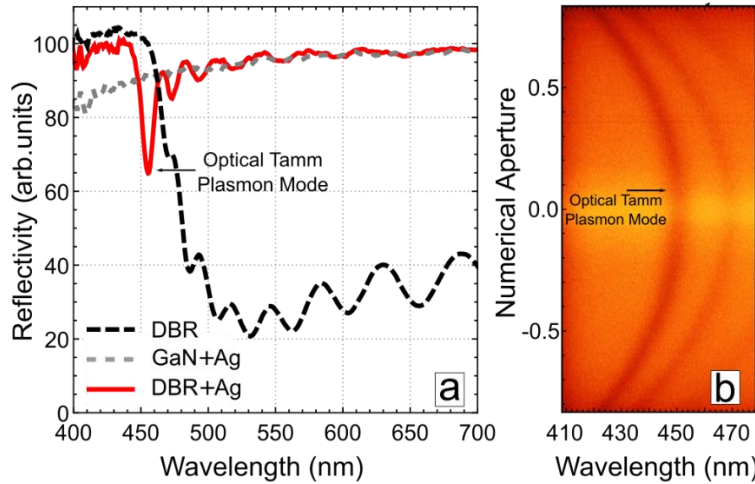

Figure 4(a) Reflectivity spectra at normal incidence of nanoporous GaN DBR (dash black line), a 55 nm Ag layer on top of a 3 μm GaN buffer on GaN substrate (grey dots) and a TP structure with a 55 nm layer of Ag on top of a nanoporous GaN DBR (solid red line). (b) Experimental dispersion relation of a 55 nm layer of Ag on top of a nanoporous GaN DBR. The parabolic TP dispersion relation of a TP structure is clearly visible.

To further evaluate the TP mode, energy-momentum reflectance spectroscopy was used to evaluate the dispersion curves for the structure. Momentum-resolved reflectance spectra were measured using techniques described thoroughly in ref. 26 and a setup similar to the ones describes in refs. 27,28. Figure 4(b) presents the experimental dispersion relation of TP Plasmon structure in TM polarization. As previously demonstrated, the TP mode exhibits a parabolic dispersion relation, characteristic of the Fabry-Perrot like optical mode. The reflectivity spectra with and without metal as well as the dispersion relation help us to conclude unambiguously on the observation of TP modes. Our first evaluations of the quality factor of the TP mode leads to ~ 300, which is lower than those reported earlier[5] (~ 1000). Increasing the number of pairs to obtain near 100% reflectance, and improving the uniformity of the DBRs may improve the quality factor. Furthermore, the use of Ag as the metallic layer may not be ideal for wavelengths ranges near 400 nm as it's approaching the plasmon resonance of Ag[29,30], which would dramatically decrease the reflectivity and thus the quality factor of the mode.  The use of aluminium or chromium[31] instead of silver could be a way to improve the quality factor of the TP on a Gan nanoporous DBR

In summary, we demonstrate a TP mode in a metal-DBR structure using a nanoporous GaN DBR system. A strong dependence of the TP mode position on the porosity and refractive index of the porous GaN layer was observed based on calculations (shift of ~ 4 nm when $n_{por}$ changes from 1.0 to 1.3), indicating potentials for optical sensing applications. An optimum thickness of Ag (~ 55 nm) was realized from theoretical calculations and was deposited on the DBRs. The DBR + 55 nm Ag structure shows a dip in reflectivity at ~ 455 nm corresponding to a TP mode. The results from mode dispersion with energy-momentum reflectance spectroscopy measurements also supports the presence of a TP mode at the metal-nanoporous GaN DBR interface. The TP on lattice-matched nanoporous GaN DBRs have applications in active and passive optoelectronic applications such as optical sensing, resonant cavity LEDs, and lasers.

The authors acknowledge the support from the Solid State Lighting and Energy Electronics Center (SSLEEC) at University of California, Santa Barbara (UCSB). This work was supported by the National Science Foundation under Grant No. DMS-1439786 and grant from the Simons Foundation (601952, JS). A portion of this work was performed at the UCSB nanofabrication facility. Energy-momentum spectroscopy measurements and analyses were supported by the National Science Foundation (DMR-1454260 and OIA- 1538893) and by the Air Force Office of Scientific Research (Grant No. FA9550-16-1-0393).